\documentclass[12pt]{article}
\usepackage[dvips]{graphicx}
\usepackage{latexsym}
\newcommand{\beq}{\begin{equation}}
\newcommand{\eeq}{\end{equation}}
\newcommand{\Frac}[2]{\frac{\displaystyle #1}{\displaystyle #2}}

\newcommand{\Oa}{{\cal O} (\alpha_{s}^3)}
\newcommand{\Oaa}{{\cal O} (\alpha_{s}^2)}

\begin{document}
\thispagestyle{empty}
\begin{titlepage}
\begin{center}
\hfill IFIC/02$-$07 \\ 
\hfill FTUV/02$-$0211 \\
\vspace*{3.5cm} 
\begin{Large}
{\bf New contributions to heavy quark sum rules
} \\[2.25cm]
\end{Large}
{ \sc J. Portol\'es} \ and { \sc P. D. Ruiz-Femen\'\i a}\\[0.5cm]
{\it Departament de F\'\i sica Te\`orica, IFIC, Universitat de Val\`encia -
CSIC\\
 Apt. Correus 22085, E-46071 Val\`encia, Spain }\\[2.5cm]

\begin{abstract}
\noindent
We analyse new contributions to the theoretical input in heavy quark 
sum rules and we show that the general theory of singularities of 
perturbation theory amplitudes yields the method to handle these
specific features. In particular we study the inclusion of heavy quark
radiation by light quarks at ${\cal O}(\alpha_s^2)$ and of 
non--symmetric correlators at $\Oa$. Closely related, we also propose
a solution to the construction of moments of the spectral densities
at $\Oa$ where the presence of massless contributions invalidates the
standard approach. We circumvent this problem through a new definition
of the moments, providing an infrared safe and consistent procedure.

\end{abstract}
\end{center}
\vfill
\hspace*{1cm} PACS~: 11.55.Hx, 11.55.Fv, 13.65.+i, 12.38.Bx \\
\hspace*{1cm} Keywords~: Heavy quark sum rules, singularities of perturbative
amplitudes.
\eject
\end{titlepage}

\pagenumbering{arabic}

\section{Introduction}
\begin{figure}[tb]
\begin{center}
\hspace*{-0.5cm}
\includegraphics[angle=0,width=0.85\textwidth]{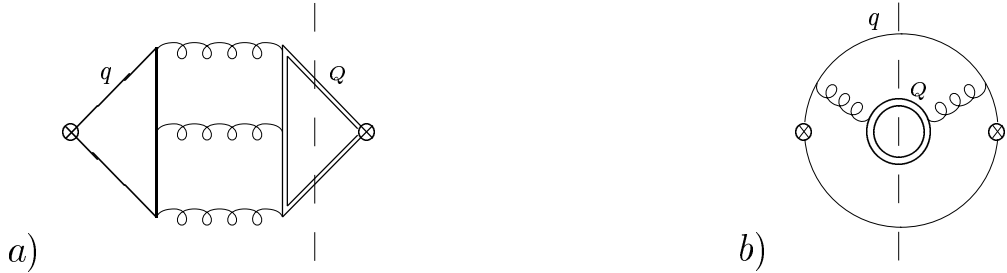}
\end{center}
\caption[]{\label{fig:5gluon} \it Examples of perturbative non--heavy
quark current correlators at ${\cal O} (\alpha_s^3)$ (a) and 
${\cal O} (\alpha_s^2)$ (b) that contribute to the production of 
$Q \overline{Q}$ states.}
\end{figure}

\hspace*{0.5cm}Sum rules analyses have extensively exploited the relation
between the correlator of the quark electromagnetic currents and the cross
section of $e^+e^- \to hadrons$ under the assumption of quark-hadron duality,
to
extract fundamental information of hadron systems. The two-point function
containing the QCD dynamics of the produced quarks is built from the sum of
the electromagnetic vector currents associated to each flavour:
\begin{eqnarray}
\Pi^{\mu \nu}_{had}(p) \; & = &  \; i \, \int \, d^4x \, e^{ipx} \,
\sum_{q,q'} \, e_q \, e_{q'} \,
 \langle \, 0 \, | \, T \, \left( \, \overline{q}(x) \, 
\gamma^{\mu} \, 
q(x) \, \right) \,
 ( \,  \overline{q'}(0) \, \gamma^{\nu} \, q'(0) \,)
 \, | \, 0 \,
\rangle \; \; \, \nonumber \\
& = & \; ( \, - \, g^{\mu \nu} \, p^2 \, + \, p^{\mu} \, p^{\nu} \, ) \, 
\Pi_{had}(p^2) \, \; , 
\label{eq:HadCorre}
\end{eqnarray}
where $q$ and $q'$ stand for heavy or light quarks, indistinctly, with 
electric charges $e_q$ and $e_{q'}$. Here we
find two
types of correlators: the symmetric ones, both electromagnetic currents
corresponding to the same flavour, and non-symmetric correlators, where 
$q \neq q'$. 
Strictly, the latter are needed to fully describe the electromagnetic
production of
hadrons, even in the case where a definite flavour type of hadrons is
isolated in the final state. Sum rules analyses applied to heavy quark
production are written
down in terms of the symmetric correlator built from the vector current
$j^\mu_{Q} (x) = e_Q \, \overline{Q}(x)\, \gamma^\mu \, Q(x)$ of 
the heavy quark $Q$,
and the effects of the non-symmetric correlators are never considered.
The reason is that they begin to contribute beyond ${\cal{O}}(\alpha_s^2)$
in QCD perturbation theory (see Fig.~\ref{fig:5gluon}(a)), which means 
one order
beyond the actual knowledge of the (symmetric) heavy quark correlator
$\Pi_{Q \overline{Q}}$. The study of such new effects in $Q\overline{Q}$
production will be mandatory if ${\cal{O}}(\alpha_s^3)$ accuracy is reached
in the future. 
However, already at ${\cal{O}}(\alpha_s^2)$ the production of heavy 
quarks $Q \overline{Q}$ receives
contributions which have neither been accounted for in the theoretical
input of heavy quark sum rules. These arise from heavy quark discontinuities
of symmetric correlators built from quarks such that $m_q < m_Q$,
as the cut shown in 
Fig.~\ref{fig:5gluon}(b), representing the production of heavy hadrons
radiated off a pair of lighter quarks. 
\par
Finally, Groote and Pivovarov have recently pointed out \cite{pg1,pg2}
that, at $\Oa$, a three--gluon intermediate state (see Fig.~\ref{fig:3gluon})
contributes to the $\Pi_{ Q \overline{Q}}$ correlator. 
As these authors have
shown, this massless intermediate state invalidates the usual definition
of the moments ${\cal M}_{n}$, 
\begin{equation}
{\cal M}_n \; = \; \Frac{1}{n!} \, \left( \, \Frac{d}{d p^2} \,
\right)^n \, \Pi_{Q \overline{Q}}(p^2) \, \bigg|_{p^2=0} \; \; ,
\end{equation}
for $n \ge 4$, when they 
become singular. Consequently the use of heavy quark sum rules at 
$\Oa$ is debatable.
\par
All the features we have just quoted arise as a consequence of the interplay
between the implementation of quark--hadron duality and the proper
definition of the observables in the case of heavy quarks QCD sum
rules. The
correlation between the perturbative input and the observable information
on the experimental side requires a careful matching that cannot be 
fully achieved. Accordingly the introduced incertitudes should be 
estimated and included in the errors of the parameters determined through
this method.
\par
Here we discuss the aspects pointed out above and their consequences in 
the methodology of extracting information from QCD sum rules. The aim 
of this work is to provide a consistent procedure to implement the
perturbative input in the theoretical side of the heavy quark sum rules. 
Our proposal relies in a careful application of the
general theory of singularities of perturbation theory. 
The crucial
point will be to isolate all the cuts related to $Q \overline{Q}$ 
production from the general vector two--point function (\ref{eq:HadCorre})
in order to construct a modified correlator containing only contributions to 
heavy quark production.
\par
In Section~2 we recall the theory of singularities of perturbative 
amplitudes. The relation between the phenomenological and the theoretical
input in the QCD sum rules is discussed in Section~3. Hence Sections~4
and 5 collect the implementation of our proposal to include heavy quark
radiation off light quarks and to exclude massless singularities,
respectively. We will comment on the uncertainties related with our
method too. In Section~6 we emphasize our conclusions.

\section{Analyticity of $\Pi_{had}(s)$}
\hspace*{0.5cm}
As it is well known two--point functions are analytic except for 
singularities at simple poles or branch cuts, the latter being originated
by normal thresholds of production of internal on--shell states.
Assuming that the absorptive part of $\Pi_{had}(p^2)$
starts at some point $s_0$, vanishing below this point,
the correlator satisfies the dispersion relation 
\cite{dera} \footnote{Sometimes the Adler function defined
as $\partial \Pi(p^2) / \partial \ln p^2$, to get rid of the 
subtraction constant, is used. The choice of the regularization
prescription is not relevant for our discussion here.}~:
\begin{equation}
\widehat{\Pi}_{had}(p^2) \, \doteq \, 
\Pi_{had}(p^2) \, - \, \Pi_{had}(0) \, = \, \frac{p^2}{\pi}\int^{\infty}_{s_0}
\Frac{ds}{s}\,\,
\frac{\mbox{Im}\,\Pi_{had}(s)}{s-p^2-i\epsilon}
\; \; .
\label{eq:disp-rel}
\end{equation}
The absorptive part $\mbox{Im}\,\Pi_{had}(s)$ is a physical observable, as it
is proportional to the total hadron production cross section by a vector
current $J^{\mu}=\sum_q j_q^{\mu}$. Being QCD the underlying theory
of strong interactions, the quark--hadron duality hypothesis allows us
to identify, inclusively, the states in terms of observable hadrons with
the partonic intermediate states. Hence the optical theorem tells us that the 
total absorptive part is the
sum of the absorptive parts corresponding to different intermediate 
partonic states:
\begin{equation}
\mbox{Im} \, \Pi_{had} (s)
 \; = \; - \, \Frac{1}{6s}\int \, \sum_n \,d R_{n} \,
\langle \, 0 \, | \, J^{\mu} \, | \, n \, \rangle \, \langle \, n
 \, | \, J_{\mu}^{ \dagger} \, | \, 0 \, \rangle 
 \; = \; \sum_n \, \mbox{Im} \, \Pi_n (s)\; \; ,
\label{eq:unitazi}
\end{equation} 
where the phase space integration has been explicitly stated
\footnote{We use $d R_{n} \, = \, (2 \pi)^4 \, \delta^4 (q - \sum_{i=1}^n p_i)
\, \prod_{i=1}^n d p_i$, where $q$ is the current four--momentum
and $d p_i = \frac{d^3 p_i}{(2 \pi)^3 2 E_i}$. The $-1/(6s)$ factor in 
Eq.~(\ref{eq:unitazi}) originates from $\Pi_{had} = - g_{\mu \nu} \,
\Pi^{\mu \nu}_{had} /( 3 s )$ and the (1/2) factor from the unitarity
relation.}.
A similar separation between contributions of different final 
hadron states in 
the perturbative evaluation of the 
two-point correlator, Eq.~(\ref{eq:HadCorre}), would allow us 
to keep only the desired heavy quark cuts in the symmetric and
non-symmetric correlators. Although Cutkosky rules provide a method to isolate
cuts corresponding to different intermediate states at the perturbative
level, some care is needed in their application.
\par
The study
of analytic properties of perturbation theory amplitudes shows that their
singularities are isolated and, therefore, we can discuss each singularity
of a perturbative amplitude by itself \cite{oldies}. As a consequence,
any one--variable dependent amplitude
$F(z)$ satisfies a dispersion relation from Cauchy's 
theorem given by
\footnote{This expression also gives the residue $R_i$ of a pole at 
$z = z_i$ if we interpret the discontinuity as 
$ \left[  \, F(z) \, \right]_n \, = \, 2\pi i R_i \delta(z-z_i)$.}~:
\begin{equation}
F(z) \, = \, \Frac{1}{2\pi i} \oint dz^{\prime}\frac{F(z^{\prime})}
{z^{\prime}-z}
\,= \, \sum_n \int_{z_n}^{\infty} \, 
\Frac{dz^{\prime}}{2\pi i} \, 
\Frac{\left[  \, F(z^{\prime}) \, \right]_n}{z^{\prime} \, - z} \; \; \; ,
\label{eq:Fz}
\end{equation}
where $\left[  \, F(z) \, \right]_n$ is the discontinuity
across a branch cut which
starts at the point $z_n$ and it is associated to a definite
intermediate state. For the
general two-point function in Eq.~(\ref{eq:HadCorre}), which depends 
on the total
momentum squared $p^2$, we would have
\begin{equation}
\widehat{\Pi}_{had} (p^2) \, = \, \sum_n \,\Frac{p^2}{2 \pi i} \, 
 \int_{s_n}^{\infty} \, 
\Frac{ds}{s} \, \Frac{\left[  \, \Pi(s) \, \right]_n}{s \, - \, 
p^2 \, - \, i\epsilon} \; \; \; ,
\label{eq:piq2}
\end{equation}
where now
$\left[ \Pi(s) \right]_n$  provides the sum of all the cut
diagrams associated 
to a definite intermediate state labeled $n$, 
($n=q\bar{q},q^{\prime}\bar{q}^{\prime},
ggg,q\bar{q}q^{\prime}\bar{q}^{\prime},\dots$). 
In the perturbative calculation, every 
discontinuity contributing to $\left[ \Pi(s) \right]_n$ can be associated to a
\lq \lq reduced" Feynman
diagram obtained by contracting internal off--shell propagators to a point
and leaving internal on--shell lines untouched.
Its contribution is written down following the Cutkosky rules for the
graph. 
However the discontinuity across
a specified cut in a single diagram needs not to be a pure real function in
the physical region. Hence the separation between the imaginary parts coming
from different final states, as stated in Eq.~(\ref{eq:unitazi}), 
does not seem to apply for individual diagrams. But from 
Eqs.~(\ref{eq:unitazi}) and (\ref{eq:piq2}) we can conclude that
$\left[  \, \Pi(s) \, \right]_n = \, 2 i \, \mbox{Im} \, \Pi_n (s)$, 
meaning that only
the sum of all cuts corresponding to a defined intermediate state provides
the physical observable, i.e.  $\mbox{Im} \, \Pi_n (s)$. Evidently, this 
holds at any perturbative order in $\alpha_s$, and gives a prescription
to isolate contributions to different quark intermediate states in the
hadron two--point function. This assertion might seem obvious but it is not~:
A $Q \overline{Q}$ cut on the right--hand fermion 
loop in Fig.~\ref{fig:3gluon}(a) does not provide,
by itself, a pure real contribution. Only
when both $Q \overline{Q}$ cuts, on the left--hand and right--hand fermion
loops of Fig.~\ref{fig:3gluon}(a), are added we get 
a term contributing to the physical observable 
$\mbox{Im} \,  \Pi_{n= Q \overline{Q}}$ .
\par
This last example also shows that some subsets of discontinuities of the same
intermediate state give already real functions prior to the summation 
of all contributions at a
fixed perturbative order. This is the case for the set of cuts coming
from a symmetric correlator, and for the set
arising from a non--symmetric correlator with currents 
$j_q^{\mu}, \, j_{q^{\prime}}^{\mu}$ 
together with
its conjugate. This is easily seen if we rewrite 
the absorptive part
corresponding to the state $n$, $\mbox{Im} \, \Pi_n$, 
as a sum of terms arising from symmetric and from non--symmetric correlators:
\begin{eqnarray}
\mbox{Im} \, \Pi_n (s)
  =  - \, \Frac{1}{6s}\int \, d R_{n}\!\!\!\!\! &\Bigg[  & \!\!\!\!\!\sum_q
\langle \, 0 \, | \, j^{\mu}_q \, | \, n \, \rangle \, \langle \, n
 \, | \, j_{q,\mu}^{ \dagger} \, | \, 0 \, \rangle
 \nonumber\\[3mm]
 \!\!\!\!\! & + & \!\!\!\!\!\sum_{m_q < m_{q^{\prime}}} \, \left( \, 
 \langle \, 0 \, | \, j^{\mu}_q \, | \, n \, \rangle \, \langle \, n
 \, | \, j_{q^{\prime},\mu}^{ \dagger} \, | \, 0 \, \rangle
 \, + \, \langle \, 0 \, | \, j^{\mu}_{q^{\prime}} \, | \, n \, \rangle \, 
 \langle \, n
 \, | \, j_{q,\mu}^{ \dagger} \, | \, 0 \, \rangle \, \right) \, \Bigg]
  .
\label{eq:Pin}
\end{eqnarray} 
The first term in the r.h.s. of Eq.~(\ref{eq:Pin}) represents the absorptive
contribution from symmetric correlators, and the perturbative expansion 
of each one, following Cutkosky rules, is clearly real. In the case of
interest, $n\equiv [Q\overline{Q}]$ \footnote{Brackets $[Q \overline{Q}]$
are short for any hadron state containing at least a 
$Q \overline{Q}$ pair and, possibly, light quarks and gluons too.}, 
this term contains the usual 
heavy quark spectral density built from heavy quark currents, 
$\Pi_{Q \overline{Q}}$, and $[Q \overline{Q}]$ production through light
quark currents correlators. The
second and third terms in Eq.~(\ref{eq:Pin}) are conjugate to 
each other, so their
sum also gives a pure real number. In terms of diagrams, this means that 
to extract the desired absorptive part from non-symmetric correlators we 
need to
add to the cut of a diagram the corresponding one in the
conjugated diagram  (see Fig.~\ref{fig:5gluon}(a); the discontinuity obtained
from the same
diagram with quark $q$ and quark $Q$ lines interchanged should be added up to
get a real contribution). 

\section{Phenomenology vs theoretical input in heavy quark sum rules}
\hspace*{0.5cm}
The analysis above shows that a clear control can be enforced on the
perturbative side of the sum rules in order to include or exclude 
specific contributions. However while
there is no doubt about the observable that provides
$\mbox{Im} \, \Pi_{had} \, \propto \, \sigma ( e^+ e^- \rightarrow hadrons)$
when an exclusive hadron sector (like, for example, heavy quark production)
is specified, it is clear that the matching between the perturbative
and the phenomenological side includes incertitudes related with the
content and definition of the final state. 
\par
Heavy quark sum rules \cite{rev} have been successful in providing information
on the heavy quark parameters. In short they make use of global 
quark--hadron duality that translates into the ansatz on the 
vector correlator $\Pi_{[Q \overline{Q}]}(s)$~:
\begin{equation}
\label{eq:gqhd}
\int_{s_0}^{\infty} \, ds \, 
\Frac{\mbox{Im} \, \Pi_{[Q \overline{Q}]}^{phen}(s)}{s^n}
\; \simeq \; 
\int_{4 M^2}^{\infty} \, ds \, 
\Frac{\mbox{Im} \, \Pi_{[Q \overline{Q}]}^{pert}(s)}{s^n} \; + \; ... \, ,
\end{equation}
where $\mbox{Im} \, \Pi_{[Q \overline{Q}]}^{phen}(s)$ on the l.h.s. gives the
phenomenological information on heavy quark production and it is
related with the cross--section of vector current production of hadrons
containing Q--flavoured states. On the
r.h.s. $\mbox{Im} \, \Pi_{[Q \overline{Q}]}^{pert}(s)$ is the QCD perturbative
contribution to the correlator, and in the lower limit of integration
$M$ is usually taken as the pole mass of the heavy quark. Finally 
the dots on the r.h.s. are 
short for non--perturbative  (the gluon condensate essentially) 
contributions and possible Coulomb--like bound states coming
from non--relativistic resummations in $\Pi_{[Q \overline{Q}]}^{pert}$ 
below threshold. These last two
features are not relevant for the discussion of this paper and have to 
be implemented on our results without modification.
\par
To a definite perturbative order in $\alpha_s$,
$\mbox{Im} \, \Pi_{[Q \overline{Q}]}^{pert}(s)$ includes all the absorptive
contributions to the correlator that provide $[Q \overline{Q}]$ production.
Notice that this is not the same that the absorptive $Q \overline{Q}$
contribution of the heavy--quark current correlator 
$\Pi_{Q \overline{Q}}$, as it is usually 
assumed. The total experimental cross section 
$\sigma (e^+ e^- \rightarrow hadrons)$ can be split into two disjoint
quantities~: the cross section for producing hadrons with
Q--flavoured states, and the production of hadrons with no Q--flavoured
components. If the experimental set up was accurate
enough to classify events into one of these two clusters, the first
class would be the required ingredient for the phenomenological part
of the heavy quark sum rule. However this separation, implemented in 
the theoretical
side within perturbative QCD, is rather involved. Up to $\Oaa$ there
has not been any doubt, in the literature, that contributions to this
side arise wholly from $Q \overline{Q}$ cuts in the heavy quark 
correlator $\Pi_{Q \overline{Q}}$. 
The physical picture behind this assertion relies in the assumption
of factorization between hard and soft regions in the quark production
process and subsequent hadronization. The hard region described
with perturbative QCD entails the production of the pair of heavy
quarks, and the soft part of the interaction is responsible for the observed
final hadron content. Although possible, annihilation of the partonic
state $Q \overline{Q}$ due to the later interaction is very unlikely,
as jets arising from the short distance interaction fly apart before
long--distance effects become essential. Consequently, each jet
hadronizes to a content of Q--flavoured states with unit probability.
As local duality is implicitly
invoked, this picture is assumed to hold at sufficient high energies; hence
perturbative corrections to the hard part are successively included
through the heavy quark currents correlator. We claim, though, that similar
$Q \overline{Q}$ cuts are present in non--symmetric correlators, 
starting at $\Oa$, as the one shown in Fig.~\ref{fig:5gluon}(a), where
the left hand part of the cut diagram is a genuine production of 
$Q \overline{Q}$ states triggered by virtual light quarks.
If the use of heavy quark sum rules
up to this order is considered, the inclusion of these terms of the
correlator of a 
heavy and a light quark currents should be taken into account. 
According to our conclusion in the last Section, once the discontinuity 
provided by Fig.~\ref{fig:5gluon}(a) is known, it has to be added to
$\mbox{Im} \, \Pi_{[Q \overline{Q}]}^{pert}(s)$. 
\par
Other extra $Q \overline{Q}$ cuts, i.e. not contained in 
$\Pi_{Q \overline{Q}}$, arise even at $\Oaa$ as the diagram of
Fig.~\ref{fig:5gluon}(b). In this case the $Q \overline{Q}$ pair is
produced through the splitting of a hard gluon radiated off
a pair of light quarks. Whether this cut should be accounted for or not
in the theoretical side depends crucially on which is the content and
the configuration of the reconstructed final state in the experimental data,
as the physical picture outlined above for pure $Q \overline{Q}$ cuts
does not apply so clearly for $Q \overline{Q} q \overline{q}$ 
discontinuities. We will come back to this point at the end of Section~4. 
In addition a discussion
about other possible contributing cuts should arise. The case of the
three--gluon discontinuity is postponed to Section~5.
\par
In the following
we will discuss, in turn, the inclusion of heavy quark radiation by light
quarks and the infrared massless discontinuities noticed by 
Groote and Pivovarov. We will provide specific solutions
along the lines put forward in Sections~2 and 3.

\section{Heavy quark radiation}
\hspace*{0.5cm}
Starting at  
${\cal O}(\alpha_s^2)$, symmetric correlators built from light 
quark currents include four 
fermion cuts with a heavy quark pair radiated off the light quarks
as shown in Fig.~\ref{fig:5gluon}(b) (two additional diagrams, one
with the two gluons attached to the lower light fermion line, and
the other with one gluon attached to each light fermion line, should
be considered too).
The sum of all these four fermion
absorptive parts in the three-loop diagrams with massless light quarks 
currents has been calculated in Ref.~\cite{4fermion}, and can be cast into the
following form \footnote{Notice that our definition of 
$R_{q\overline{q}Q\overline{Q}}$ differs from the one in 
Ref.~\cite{4fermion}.}~:
\begin{equation}
12\pi \, \mbox{Im} \, \Pi_{q\overline{q}Q\overline{Q}} (s) \, = \, 
R_{q\overline{q}Q\overline{Q}} \, \equiv \, N_c \,
\big( \!\! \sum_{i=u,d,s} Q_i^2 \big)  
\, C_{8} \, \bigg( \frac{\alpha_s}{\pi} \bigg)^2 \int_{4M^2}^s 
\frac{ds^{\prime}}{s^{\prime}} \,
R(s^{\prime}) \, F(s^{\prime}/s) \; \; \; ,
\label{eq:qqQQ}
\end{equation}
with 
\begin{eqnarray}
F(x) &=&  \frac{1}{6} \, \bigg\{ (1+x)^2\ln^2 x + (3+4x+3x^2)\ln x
+5(1-x^2) \nonumber\\[3mm]
&&\; \; \, \; \; \; - 4(1+x)^2 \, 
\Big[ \mbox{Li}_2(-x) +\ln(1+x)\ln x + \Frac{\pi^2}{12} \,\Big]
\bigg\} \; \; \; . 
\label{eq:F}  
\end{eqnarray}
The function $F(s^{\prime}/s)$ gives the rate for the decay of a vector boson
of mass $\sqrt{s}$ into a vector boson of mass 
$\sqrt{s^{\prime}}$ plus a pair of massless
fermions ($q\overline{q}$). The spectral density 
$R(s) = \beta (3 - \beta^2)/2$ (at lowest order) is the normalized 
cross section for the production of a pair of fermions with unit charge 
through a vector boson; here  $\beta = \sqrt{1-4M^2/s}$ is the
velocity of the produced heavy quarks. The integral can be solved 
analytically in this case
and the result is found in Ref.~\cite{4fermion}.
Note that the heavy quark pair is created in a colour
octet state, and the factor 
$$
C_{8}= \frac{1}{N_c} \,
\mbox{Tr}\bigg(\frac{\lambda^a}{2} \frac{\lambda^b}{2}\bigg) \,
\mbox{Tr}\bigg(\frac{\lambda^a}{2} \frac{\lambda^b}{2}\bigg) \, 
= \, \frac{2}{3}
$$ 
retains this colour structure. It is interesting to compare
the contribution from $R_{q\overline{q}Q\overline{Q}}$ with the 
${\cal O}(\alpha_s^2)$
contributions to $R_{Q\overline{Q}}$ (i.e. to the spectral density 
of the heavy quark correlator). Note that in the high energy limit there is no
difference between the diagram shown in Fig.~\ref{fig:5gluon}(b) and the same
one with $Q$ and $q$ lines interchanged or with $q=Q$, both of them 
being included in $\Pi_{Q\bar{Q}}$. Differences arise because the
heavy quark currents correlator,
$\Pi_{Q\overline{Q}}$, also accounts for two heavy quark
cuts where the internal (light or
heavy) quark loop represents a virtual correction to the electromagnetic
current. 
\par
We have written Eq.~(\ref{eq:qqQQ}) in terms of a general
$R(s)$ function in the integrand because it allow us to
introduce in a straightforward way final state interactions between the heavy
quark pair. In particular, we know that close to threshold the Coulomb
interaction between the heavy quark pair dominates the dynamics. Resummation of
leading terms $\sim (\alpha_s/\beta)^n$ becomes mandatory, 
and gives rise to the
well known Sommerfeld factor multiplying the cross section:
\begin{equation}
R^{thr}(s)=R(s)\times\frac{C\pi \alpha_s / \beta}{1-\exp
(-C\pi\alpha_s/\beta)}\; \; \; . 
\label{eq:Rthr} 
\end{equation}
The colour factor $C$ appears in the Coulomb QCD potential and its value depends
on the relative colour state of the quark pair. 
For singlet states $C = C_F$, and
the potential is attractive, increasing the cross section at threshold. This
is the case of heavy quark production in $e^+e^-$ collisions. 
However, in our case the heavy quark pair is produced through the splitting of a
gluon. The Coulomb potential becomes 
repulsive between quarks in a colour octet state, $C = C_F - C_A/2 = -1/(2N_c)$,
and the Sommerfeld factor at low velocities then reads 
$$
\frac{-\pi \alpha_s /6 \beta}{1-\exp
(\pi\alpha_s/6\beta)}\; \; \stackrel{\beta\to 0}{\Longrightarrow}\; \;
\frac{\pi \alpha_s}{6 \beta}\, e^{-\frac{\pi\alpha_s}{6\beta}}\; \; \; , 
$$
causing the cross section to 
decrease near threshold even faster than $\beta$, the phase-space velocity in
$R(s)$. The production of heavy quarks radiated off massless quarks through
a virtual gluon is then very much suppressed in the threshold region. 
However, as mentioned above, high energy quark lines can be considered
massless and the contribution from this diagram is numerically equal
to the same one with $Q$ and $q$ lines interchanged.
\par
The inclusion in 
$\mbox{Im} \Pi_{[Q \overline{Q}]}^{pert}(s)$ of four--fermion cuts coming
from light--quark correlators is possible because we have shown in Section~2
how to discern and extract these pieces. As discussed before, the procedure
depends crucially on the definition of the observable information input
in the sum rule, and consistence between the theoretical and phenomenological
parts is required. Let us come back to the discussion of Section~3. There
it was argued why perturbative $Q \overline{Q}$ cuts are thought to 
reproduce the phenomenology of two jet events. Notice that, 
in heavy quark radiation from light quarks, the signature of the event
is likely to be a 3--jet configuration where one of the jets is
generated from a gluon. If heavy flavour components are to be found
in this jet, the diagram of Fig.~\ref{fig:5gluon}(b) would certainly
be needed to account for these events in the theoretical side. However
the heavy partons in this jet are not as energetic as in a pure
$Q \overline{Q}$ production and, consequently, the proposed factorization
between long and short distance effects may not longer apply, allowing
for an interference between both regimes. In this case we cannot argue
that these kind of cut diagrams would result in a final state
with Q--flavoured hadrons with unit probability, although we may impose
kinematical constraints to reduce uncertainties in both the experimental
reconstruction of data and theoretical cross section of these
$Q \overline{Q} q \overline{q}$ states. This issue is the source of a 
recent discussion in the literature related with the secondary
production of $b \overline{b}$ through gluon splitting 
\cite{4fermion,gls1}.

\section{Massless contribution to heavy quark sum rules}
\begin{figure}[tb]
\begin{center}
\hspace*{-0.5cm}
\includegraphics[angle=0,width=0.7\textwidth]{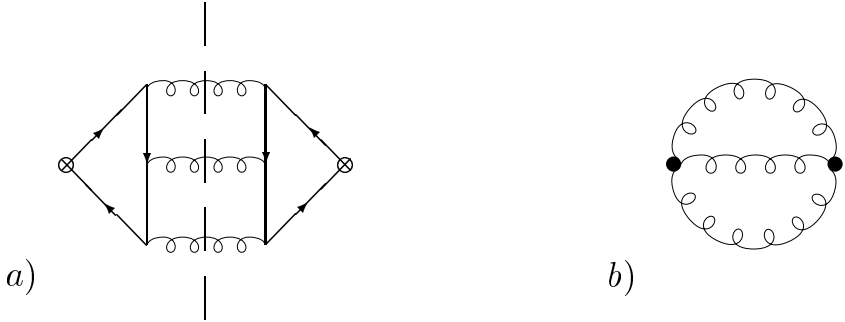}
\end{center}
\caption[]{\label{fig:3gluon} \it (a) $\Oa$ diagram contributing to the
vacuum polarization function of the heavy quark current (the vertical 
dashed line indicates the massless cut). (b) \lq \lq Effective" diagram
obtained by integrating out the fermion loops. It also has the topological
structure of the \lq \lq reduced" diagram that determines the massless
cut singularity.}
\end{figure}

\hspace*{0.5cm}
Until present the evaluation of the perturbative two--point correlation
function $\Pi^{pert}(q^2)$ (in this Section
we will denote the heavy quark currents correlator by $\Pi(q^2)$) has only 
been carried out completely, with 
massive quarks, up to ${\cal O}(\alpha_s^2)$ \cite{cher} and the
sum rules procedure, given by Eq.~(\ref{eq:gqhd}), has been termed
consistent and effective in its task because the first branch point
is set at the massive two--particle threshold. 
However Groote and Pivovarov have pointed out \cite{pg1} that at 
$\Oa$ there is a contribution to the
correlator which contains a three--gluon massless intermediate state 
(see Fig.~\ref{fig:3gluon}(a)).
Its absorptive part starts at zero energy and, therefore, 
Eq.~(\ref{eq:gqhd}) is no longer correct because on the r.h.s. there 
is a discontinuity starting at $s=0$. Moreover those authors have
also warned about the fact that, at this perturbative order, the massless
intermediate state invalidates the definition of the moments ${\cal M}_n$
for $n \ge 4$ because they become singular. 
Let us collect their reasoning.
\par
The perturbative contribution given by the diagram in 
Fig.~\ref{fig:3gluon}(a)
has been calculated at small $q^2$ ($q^2 \ll M^2$) in 
Ref.~\cite{pg1}. In this limit the quark triangle loop can be integrated
out and it ends up in the diagram in Fig.~\ref{fig:3gluon}(b) generated
by an induced
effective current describing the interaction of the vector current
with three gluons,
\begin{equation} 
J^{\mu}\,=\, - \Frac{\pi}{180 M^4} \, \left( \Frac{\alpha_s}{\pi} 
\right)^{\frac{3}{2}} \, \left(5\,
\partial_{\nu}{\cal
O}_{1}^{\mu\nu}\,+\,14\,\partial_{\nu}{\cal O}_{2}^{\mu\nu}\right)
\,,
\label{eq:effec_current}
\end{equation}

\noindent
with
\begin{eqnarray} 
{\cal O}_{1}^{\mu\nu} \,  & = & \, d_{abc}\,G^{\mu\nu}_a G^{\alpha\beta}_b 
G_{\alpha\beta}^c \; \; , \\ \nonumber
{\cal O}_{2}^{\mu\nu} \, & = & \, d_{abc}\,G^{\mu\alpha}_a G_{\alpha\beta}^b 
G^{\beta\nu}_c \; \; ,
\label{eq:operators} 
\end{eqnarray}  
where $G^{\mu\nu}_a$ is the gluon strength field tensor. The effective 
current in the QED case 
($G^{\mu\nu}_a\to F^{\mu\nu}, \alpha_s \to \alpha_{em}, d_{abc}\to 1$) can be 
easily identified from the
lowest order Euler-Heisenberg Lagrangian (see Ref.~\cite{pg2}).  
\par
The correlator of the induced current (\ref{eq:effec_current}) is then
evaluated in the configuration space giving~:
\begin{equation}
\langle 0| T \,J_{\mu}(x) \
J_{\nu}^{\dagger}(0)\, |0\rangle\,=\,-\frac{34}{2025\pi^4 M^8}
\left(\frac{\alpha_s}{\pi}\right)^3 d_{abc}d_{abc}\,
\left(\partial_{\mu}\partial_{\nu}-g_{\mu\nu}\partial^2\right)
\frac{1}{x^{12}} \, .
\label{eq:correlator}
\end{equation}
In momentum space we need to perform the Fourier transform of
Eq.~(\ref{eq:correlator}). Following the
differential regularization procedure \cite{differ}, which works
directly in configuration space, the result for the vacuum polarization
contribution of the diagram in Fig.~\ref{fig:3gluon}(b) at small $q^2$ reads
\begin{equation} 
 \Pi_{\mu\nu}(q)\; = \; \frac{17}{2916000 \pi^2}\,d_{abc}d_{abc}
\left(\frac{\alpha_s}{\pi}\right)^3
(q_{\mu}q_{\nu}-q^2g_{\mu\nu})\left(\frac{q^2}{4M^2}\right)^4
\ln \left(\frac{\mu^2}{-q^2}\right)\,
+{\cal O}\Big[ \Big(\frac{q^2}{M^2}\Big)^5 \Big]\,,
\label{eq:3gluon_polarization}
\end{equation}
with $\mu$ the renormalization point in this scheme,
and $d_{abc}d_{abc}=40/3$. 
\par
As noticed by Groote and Pivovarov \cite{pg1}, moments associated to the
diagram in Fig.~\ref{fig:3gluon}(b) are not defined if $n\ge 4$. 
Indeed differentiating 
Eq.~(\ref{eq:3gluon_polarization}) four times, at $q^2\approx 0$, we get:
\begin{equation}
\frac{1}{4!}\left(\frac{d}{dq^2}\right)^4\Pi(q^2)\arrowvert_{q^2\approx 0}=
\frac{17}{218700\pi^2}\left(\frac{\alpha_s}{\pi}\right)^3
\left(\frac{1}{4M^2}\right)^4
\left[\ln \left(\frac{\mu^2}{-q^2}\right)-\frac{25}{12}\right]
+{\cal O}\Big[ \frac{q^2}{M^{10}} \Big]
\; \; ,
\label{eq:moment4}
\end{equation}
whose real part clearly diverges if we set $q^2=0$. 
Larger $n$ moments are also
infrared divergent, and so the authors of Ref.~\cite{pg1} conclude
that the standard sum rule analysis must limit the accuracy of theoretical
calculations for the $n\ge 4$ moments to the ${\cal O}(\alpha_s^2)$ order of
perturbation theory. This is, essentially, the conclusion of Ref.~\cite{pg1}.
\par
An
infrared safe redefinition of the moments, to cure the latter problem,
has been provided in Ref.~\cite{pg2}; it consists in evaluating the 
moments at an 
Euclidean point $q^2 = - s_E$, \mbox{$s_E > 0$}, 
thus avoiding the singular behaviour.
This solution, as explained by the authors of that reference, is rather 
ill--conditioned from the phenomenological side though.
Nevertheless the fault in Eq.~(\ref{eq:gqhd}) due to the massless
threshold still represents a problem because even if, up to $\Oa$, 
we substitute the dispersion relation by 
\begin{equation} 
\widehat{\Pi}^{pert}(q^2) \;  = \; 
\Frac{q^2}{\pi}\int^{\infty}_{4M^2} \, 
\Frac{ds}{s} \, \, 
\frac{\mbox{Im}\,
{\Pi}_{ Q \overline{Q}}^{pert}(s)}{s-q^2-i\epsilon}\; + \; 
\Frac{q^2}{\pi}\int^{\infty}_{0} \, \Frac{ds}{s} \,\,
\frac{\mbox{Im}\,\Pi_{3g}(s)}{s-q^2-i\epsilon}
\; \; \; ,
\label{eq:fulldisp-rel}
\end{equation}
(where $\mbox{Im}\,{\Pi}_{ Q \overline{Q}}^{pert}(s)$
includes discontinuities
starting at $s=4M^2$), 
the spectral
function $\mbox{Im} \, \Pi_{3g}(s)$ associated to the cut in 
Fig.~\ref{fig:3gluon}(a) would hardly be implemented phenomenologically 
as gluons hadronize to both heavy and light quark pairs. 
We wish to provide a bypass to recover the balance between the
right-hand and left-hand parts of Eq.~(\ref{eq:fulldisp-rel}). We will
now see that
if one does not insist in using full 
vacuum polarization for the sum rule analysis there is a way to 
overcome this infrared problem.
\par
In the heavy quark correlator the discontinuity across the 
three--gluon cut
gives a contribution to the spectral function that is unequivocally 
real~:
\begin{equation}
\Frac{1}{2i} \, \left[ \, \Pi(s) \, \right]_{3g} \, = \,  
\mbox{Im} \, \Pi_{3g} (s)
 \; = \; - \, \Frac{1}{6s}\int \, d R_{3g} \,
\langle \, 0 \, | \, j^{\mu} \, | \, 3 \, g \, \rangle \, \langle \, 3 \, 
g \, | \, j_{\mu}^{ \dagger} \, | \, 0 \, \rangle  \; \; ,
\label{eq:unita}
\end{equation}
from which the dispersive part can be evaluated independently of the
$Q \overline{Q}$ cuts. 
Accordingly we conclude that we can identify and isolate the troublesome
massless cut contribution to the two--point function. Indeed 
Eqs.~(\ref{eq:piq2}) and (\ref{eq:unita}) justify our previous
Eq.~(\ref{eq:fulldisp-rel}).
\par
Let us go back then to Eq.~(\ref{eq:fulldisp-rel}). All the difficulty
with the phenomenological application of the sum rules is now the fact that
the contribution from the three--gluon cut is contained in both sides of
the equality. 
This intermediate state hadronizes completely into hadrons with a 
content of light and/or heavy quarks indistinctly.
It is conspicuous that if we could disentangle the heavy quark hadronization,
$3g \rightarrow Q \overline{Q}$, we should include only this piece into
the sum rule. Then the singularity at $q^2 = 0$ would disappear because 
heavy quarks are produced starting at $q^2 = 4M^2$. However there is no
way to sort out light and heavy quark production off three gluons and,
therefore, if we extract
this contribution from the heavy quark sum rules we are introducing
an incertitude in the procedure because we make sure that there is no 
light quark hadronization but we miss the heavy quark production. 
It is easy to see that the induced error is small, due to the fact that
three gluons hadronize mostly to light hadrons. On one side, in the very
high energy region and following perturbative QCD with $N_F = 4$,
we have only a $1/4 \, = \, 25 \, \%$ probability of finding
a specified pair of heavy quarks produced. And this is a generous upper limit
because when we go down in energy, phase space restrictions severely reduce
the counting of heavy quarks. Hence we estimate that excluding
the three--gluon cut we introduce a tiny very few percent error in the
sum rules procedure. 
\par 
Thus we propose an {\it infrared safe}
definition of the moments by the trivial subtraction~:
\begin{eqnarray}
\widetilde{{\Pi}}^{pert} (q^2)& \doteq & \, 
\widehat{\Pi}^{pert}
(q^2) \, -
\Frac{q^2}{\pi}\int^{\infty}_{0} \, \Frac{ds}{s} \,
\frac{\mbox{Im}\,\Pi_{3g}(s)}{s-q^2-i\epsilon} \; = \; 
\Frac{q^2}{\pi}\int^{\infty}_{4M^2} \, \Frac{ds}{s} \,
\frac{\mbox{Im}\,
{\Pi}_{ Q \overline{Q}}^{pert} (s)}{s-q^2-i\epsilon}
\; \; ,
\label{eq:safe_def_pol}\\[5mm]
\widetilde{\cal M}_n & \doteq & 
{\cal M}_n-
\frac{1}{\pi}\int^{\infty}_{0}ds\,
\frac{\mbox{Im}\,\Pi_{3g}(s)}{s^{n+1}}\, \; \; .
\label{eq:safe_def_moments}
\end{eqnarray}
Of course Eqs.~(\ref{eq:safe_def_pol}) and ~(\ref{eq:safe_def_moments})
are meaningless unless we give a precise prescription about how to subtract the
contribution of the massless cuts represented by $\mbox{Im}\,\Pi_{3g}$.
Our previous discussion gives us the tool to proceed.
Once the full $\Oa$ $\Pi^{pert}(s)$ is calculated we can 
extract the imaginary part starting at $s=0$ (which should go
with a $\theta(s)$ function) for any value of $s$. 
It is clear that the $\theta(s)$ and $\theta(s-4M^2)$ terms in the imaginary
part of the vacuum polarization function correspond to three--gluon
massless and
to $Q\overline{Q}$ cut graphs, respectively, and $\mbox{Im}\,\Pi_{3g}$
and $\mbox{Im}\,\Pi_{ Q \overline{Q}}^{pert}$ are easy to distinguish, as 
Eq.~(\ref{eq:unita}) prevents the appearance of mixed
\mbox{$\theta(s)\cdot\theta(s-4M^2)$} terms.
Therefore we 
identify 
$\mbox{Im}\,\Pi_{3g}$ and we now plug it in the dispersion
integral of the right--hand side of Eq.~(\ref{eq:safe_def_moments}) 
and perform such
integration. Divergences contained in both this integral and 
${\cal M}_n$ as $q^2\to 0$ will cancel with each other if the same
infrared regularization is employed in the two quantities. 
The intuitive choice would be a low-energy cutoff $s_0 > 0$, and 
Eq.~(\ref{eq:safe_def_moments}) would be more precisely written as:
\begin{equation}
\widetilde{\cal M}_n  \; \equiv \;  \lim_{s_0\to 0^+}\left[
\frac{1}{n!}\left(\frac{d}{dq^2}\right)^n\Pi^{pert}(q^2)
\arrowvert_{q^2 = -s_0} \, 
- \,
\frac{1}{\pi}\int^{\infty}_{0} \, \Frac{ds}{s} \,
\frac{\mbox{Im}\,\Pi_{3g}(s)}{(s+s_0)^{n}}\right]\; \; ,
\label{eq:safe_moments_reg}
\end{equation}   
where a vanishing term in the $s_0 \rightarrow 0^+$ limit has been
omitted.
\par
The evaluation of
the ${\cal M}_n$ moments at $q^2=0 < 4 M^2$ made sense because, up to 
${\cal O}(\alpha_s^2)$, this point, being far away of the
heavy quark production threshold, is unphysical and the moments
are well defined through an analytic continuation from the high--energy
region. However note that the absorptive three--gluon contribution starts at
$q^2=0$ and perturbative QCD becomes unreliable. This introduces a further
new difficulty in evaluating ${\cal M}_n$ moments
at $q^2 = 0$, as we reach the physical non--perturbative region. Our
definition of the moments, $\widetilde{\cal M}_n$ in 
Eq.~(\ref{eq:safe_def_moments}), skips this problem by fully 
eliminating the massless terms and, therefore, the final heavy quark sum 
rule will
only involve physics at $q^2 > 4 M^2$, apart from possible bound 
states.
\par
\begin{figure}[tb]
\begin{center}
\hspace*{-0.5cm}
\includegraphics[angle=0,width=0.3\textwidth]{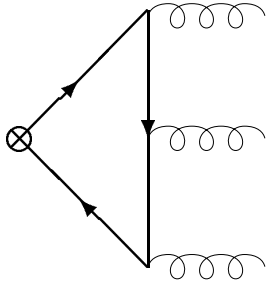}
\end{center}
\caption[]{\label{fig:4gluon} \it Feynman diagram for the production of
three gluons at ${\cal O} (\alpha_s^{3})$.}
\end{figure}

The general rule given above is valid for all orders of perturbation
theory, but it strongly relies in our ability to extract the
massless absorptive part from the full result of $\Pi(q^2)$
calculated at a definite order. Beyond ${\cal O}(\alpha_s^2)$
complete analytical results for the heavy quark correlator would be
cumbersome 
and only numerical approaches may be at hand. In this
sense, it would be convenient to have a method to
calculate $\mbox{Im}\,{\Pi}_{ Q \overline{Q}}^{pert}$ only based  
on Feynman graphs. We have already
sketched such a method in the discussion following
Eq.~(\ref{eq:piq2})~:  we just need to sum up all the massless cut
graphs to get $\mbox{Im}\,\Pi_{3g}$, and then proceed with the dispersion
integration that gives the associated dispersive part \cite{perhaps}.
For example, at 
$\Oa$, the only massless absorptive part comes from the three--gluon
cut in the diagram of Fig.~\ref{fig:3gluon}(a); let us call 
${\cal M}_{3g}^{\mu}$
the amplitude producing three gluons from the heavy quark current at
lowest order (i.e. through the quark triangle loop in
Fig.~\ref{fig:4gluon}). The massless contribution to the absorptive part
of the
correlator is then:
\begin{equation}
\mbox{Im}\,\Pi_{3g} (s) = -\frac{1}{6s}
\int dR_{3g} \,\,{\cal M}_{3g}^{\mu} \cdot {\cal M}_{3g\,\mu}^* 
\; \; ,
\label{eq:alpha3_Img}
\end{equation}  
with the three--gluon phase space integral defined as 
\begin{equation}
\int  dR_{3g} \equiv \frac{1}{3!}\frac{1}{(2\pi)^5}\frac{\pi^2}{4s}
\int_0^s ds_1 \int_0^{s-s_1} ds_2
\, \; \, ,
\label{eq:3gluon_space}
\end{equation}
in terms of the invariants $s_1\equiv (k_1+k_2)^2=(q-k_3)^2$ and
$s_2\equiv (k_2+k_3)^2=(q-k_1)^2$, and $k_i$ being the momenta of
the gluons. The real part would be obtained by integrating 
Eq.~(\ref{eq:alpha3_Img})~:
\begin{equation}
\Frac{s_0}{\pi}\int^{\infty}_{0} \, \Frac{ds}{s} \,
\frac{\mbox{Im}\,\Pi_{3g}(s)}{s+s_0} =
\Frac{-s_0}{288(2\pi)^4}\int^{\infty}_{0}
\frac{ds}{s^3(s+s_0)}\int_0^s ds_1 \int_0^{s-s_1}ds_2
\,\,{\cal M}_{3g}^{\mu} \cdot {\cal M}_{3g\,\mu}^*
\,,
\label{eq:alpha3_Reg}
\end{equation}
which, in principle, could be performed also numerically.  
The $n$th-derivative of relation (\ref{eq:alpha3_Reg}) respect 
to $s_0$, in the limit $s_0\to 0^+$, would give the infrared divergent
contribution that should be subtracted from the full moments,
as dictated by Eq.~(\ref{eq:safe_moments_reg}).  
\par
Finally, we would like to mention that using the non--relativistic
expansion of the heavy quark correlator in sum rules analyses does
not avoid this infrared problem, at least formally. 
The $\Oa$ diagram of Fig.~\ref{fig:3gluon} will be highly suppressed
in the velocity expansion, following the non--relativistic
effective field theory approach, and therefore it is not relevant
in the corresponding heavy quark currents correlator. However such two--point
function cannot describe the $Q \overline{Q}$ spectrum for energies
far from threshold and even when higher n--moments, which strongly
enhance the threshold, are used, perturbative QCD is needed
in order to implement the remaining high--energy region; the diagram
of Fig.~\ref{fig:3gluon} has to be accounted for to include 
properly this input, and its discontinuity at $s=0$ cannot be
obviated. This point is more clearly seen by noticing that, besides
the resummations in $(\alpha_s/\beta)$ performed in the non--relativistic
correlator, one could improve such expansion by adding the terms
needed to reproduce the exact $\Oa$ result $\Pi(q^2)$.

\section{Conclusions}
\hspace*{0.5cm}
Heavy quark sum rules, relying in global quark--hadron duality, are
a compelling procedure to extract information on the theory from 
phenomenology. However, as higher perturbative order analyses  are 
performed, the consistency of the method demands the inclusion of
novel features.
While at ${\cal O}(\alpha_s)$ the correlator of two heavy quark currents
gives the full perturbative information, at ${\cal O}(\alpha_s^2)$ we have
noticed that a heavy quark $Q \overline{Q}$ pair radiated from light
quarks in a correlator of light quark currents should be considered. 
At $\Oa$ the complexities grow with the essential role of non--symmetric
correlators. Closely related with this situation is the feature recently
pointed out by Groote and Pivovarov on the uneasy problem arising from 
a massless three--gluon discontinuity in the heavy quark current 
correlator at $\Oa$.
\par
We have shown that rigorous results of the general theory of singularities
of perturbation theory provide all--important tools to analyse the new
contributions. The inclusion or exclusion of specific discontinuities
in the perturbative side is shown to be feasible and the decision involves
a clear definition of the observable input on the phenomenological side of
the sum rules. 
\par
A solution for the problem pointed out by Groote and Pivovarov at 
$\Oa$ has been given. We conclude that the appropriate procedure to obtain
 information 
about the heavy quark parameters should make use of the infrared safe
corrected 
moments, defined in Eq.~(\ref{eq:safe_moments_reg}), that now indeed
satisfy the modified sum rule~:
\begin{equation}
\widetilde{\cal M}_n \; = \; \frac{1}{\pi}\int^{\infty}_{4 M^2}ds\,
\frac{\mbox{Im}\,\Pi_{[Q \overline{Q}]}^{phen}(s)}{s^{n+1}} \; \; ,
\label{eq:finali}
\end{equation}
where the right--hand side can be extracted from the heavy quark
production cross section $\sigma(e^+e^- \rightarrow [Q \overline{Q}])$.
The incertitude associated to heavy quark hadronization of the
three--gluon should be taken into account but it is shown to be tiny.
\par
The analysis we have carried out is completely general, relying in the
theory of singularities of perturbative theory amplitudes only, and 
provides a sharp tool for the future analysis of heavy quark sum rules.
\vspace*{1cm} \\
\noindent
{\large \bf Acknowledgements}\par
\vspace{0.2cm}
\noindent 
We wish to thank A.~Pich
for calling our attention on this problem.
We also thank G.~Amor\'os, M.~Eidem\"uller and A.~Pich for relevant 
discussions on the
topic of this paper and for reading the manuscript. 
The work of P.~D. Ruiz-Femen\'\i a has been partially supported by a FPU
scholarship of the Spanish {\it Ministerio de Educaci\'on y Cultura}.
This work has been supported in part by TMR, EC Contract No. 
ERB FMRX-CT98-0169 and by MCYT (Spain) under grant FPA2001-3031.

\end{document}